\documentclass[aps,prl,reprint,groupedaddress]{revtex4-2}

\usepackage[colorlinks,linkcolor=blue,citecolor=blue,urlcolor=magenta]{hyperref}
\usepackage{graphicx}
\usepackage{amsmath}

\begin{document}
	
\title{Cohesion mediated layering in sheared grains}

\author{Khushi Mahajan}
\author{Chamkor Singh}
\email[]{chamkor.singh@cup.edu.in}
%\thanks{}
\affiliation{Department of Physics, Central University of Punjab, Bathinda 151401, India}
\date{\today}

\begin{abstract}
We consider pattern formation in a sheared dense mixture of cohesive and non-cohesive grains. Our findings show that cohesive grains, which would typically form distributed agglomerates, instead segregate into percolating stripes or layers when the cohesive grain concentration ($c_o$) and cohesion strength ($C$) increase -- in a way that the average agglomerate size and the average normal stress collapse onto a single curve when plotted against $c_oC$.
Our central proposal is that the development of interfaces between cohesive and non-cohesive grains is akin to phase separation in binary molecular mixtures driven by an effective free energy, although we are dealing with a non-equilibrium system; we setup the segregation flux such that the effect of this free energy is activated only upon application of the external driving.
By constructing the segregation flux proportional to the gradient of the variational derivative of the free energy, we closely reproduce the layering in the steady-state limit.
We find a robust correspondence between the parameter $c_oC$ in the discrete simulations and the parameters in the free energy. 
\end{abstract}

\maketitle

Cohesion in granular mixtures is an important aspect for a multitude of natural and industrial settings: agglomeration in powders, mixing of seeds in soil in agriculture, wet soil aggregates in landslides and debris flows, or aggregate formation in snow avalanche deposits.
For example, by simply adding moisture as a cohesive agent, one can switch from mixing of grains to segregation of grains, utilizing differential surface wetting properties of particles~\citep{li2003controlling}, or one can alter the extent of mixing or segregation by changing the volume or viscosity of the added moisture~\citep{samadani2000segregation}.
Dry binary mixtures usually segregate due to size difference between the two species, due to differential drag or differential substrate friction acting on the two species, or due to difference between the surface or dissipative properties of the two species.
This leads to pattern formation, for example, the emergence of streak patterns due to size difference~\citep{ottino2000mixing}, band formation due to size difference~\citep{hill1995reversible} or due to differential frictional properties of the particles~\citep{newey2004band}, and stripe formation due to differential substrate friction acting on the particles~\citep{pooley2004stripe}. 
Although effect of size or particle property difference is intensively and extensively explored, little is known about pattern formation in a granular system that is binary in the sense that one of its components is cohesive. In this work, we show that cohesion mediation in conjunction with shear, which would normally lead to distributed and irregular shape agglomerates, gives rise to percolating layer or stripe formation at sufficiently increased cohesive grain concentration and cohesion strength. We predict the patterns by constructing an appropriate free energy functional and then deriving the segregation fluxes. This approach accurately describes the steady-state features of the layering patterns. Most importantly, we establish correspondence between the microscopic parameters of the discrete particle simulations and the parameters in the free energy.
\begin{figure}[t!]
	\centering
	\includegraphics[width=1.0\linewidth]{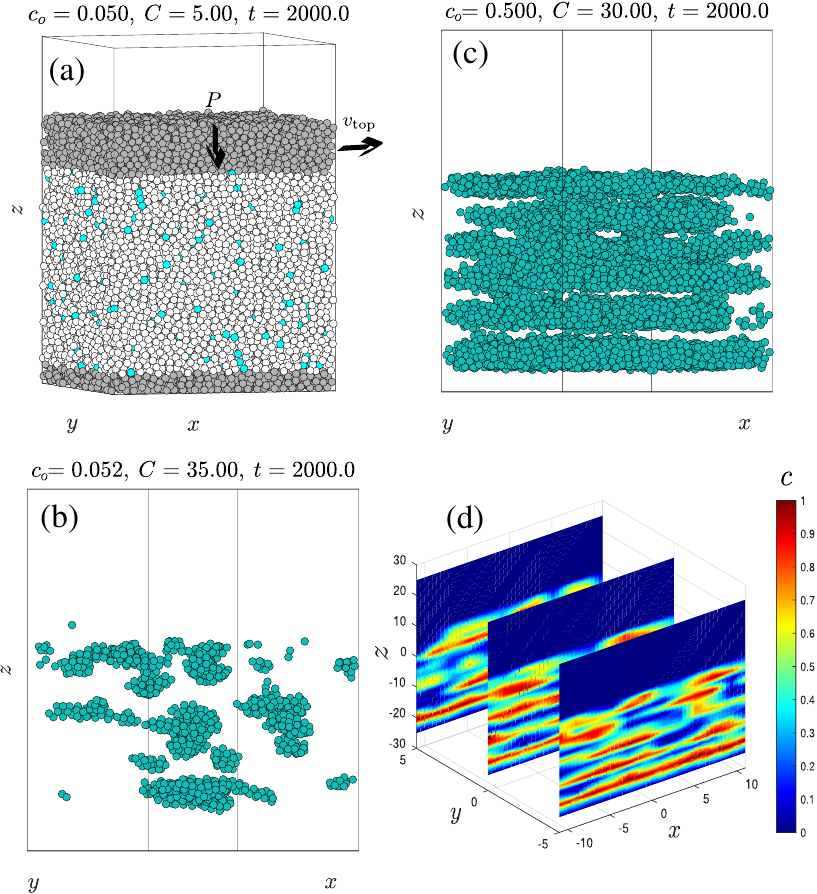}
	\caption{(a) Mixture of cohesive (green) and non-cohesive (white) grains subjected to shear. The topwall moving with a velocity $v_\mathrm{top}\mathbf{i}$ (gray) while the bottomwall remains stationary. The gravity is made to act only on the topwall particles, resulting in a pressure in the bulk. Here, $c_o$ is the overall volume fraction or concentration of cohesive grains, $C$ is a cohesive force parameter, and $t$ is the simulation time. (b) Irregular agglomerates at relatively lower $c_o$ and higher $C$, while in (c) we observe ordered layers at higher $c_o$ and $C$. Only cohesive grains are shown in (b, c). (d) Coarse-grained concentration field corresponding to (c). The coarse-graining helps us extract the average interfacial energy for comparison with the continuum model.}
	\label{fig_particles}
\end{figure}
We subject the mixture of cohesive and non-cohesive grains to plane shear using discrete element method (DEM) simulations and tune: ($i$) the overall volume fraction or concentration of the cohesive grains, $c_o$, and ($ii$) a cohesive force parameter, $C$. 
Cohesion originates due to various physical or physicochemical mechanisms: wetting or moisture absorption~\citep{mitarai2006wet}; due to surface treatment of particles with polymeric agents~\citep{gans2020cohesion}; chemical salinization~\citep{jarray2019cohesion}; or because of Van der Waals or electrostatic interactions~\citep{lee2015direct,shinbrot2018multiple,yoshimatsu2018segregation,singh2019aggregation}. It alters the dynamics and rheology in granular flows~\citep{brewster2005plug, rognon2008dense, royer2009high, mandal2020insights, vo2020additive, elekes2021expression, macaulay2021viscosity, metzger2022sinking}, and often inhibits mixing. 
Instead of considering specific nature of a cohesive interaction, we use an inter-particle cohesive force that increases linearly as particles approach each other. The parameter $C$ in our simulations represents the net interparticle attractive force in static balance measured in units of particle weight -- also known as the granular Bond number. 
The details of interactions, DEM simulations, and parameters are given in the Supplementary.

\begin{figure}
	\centering			
	\includegraphics[width=1.0\linewidth]{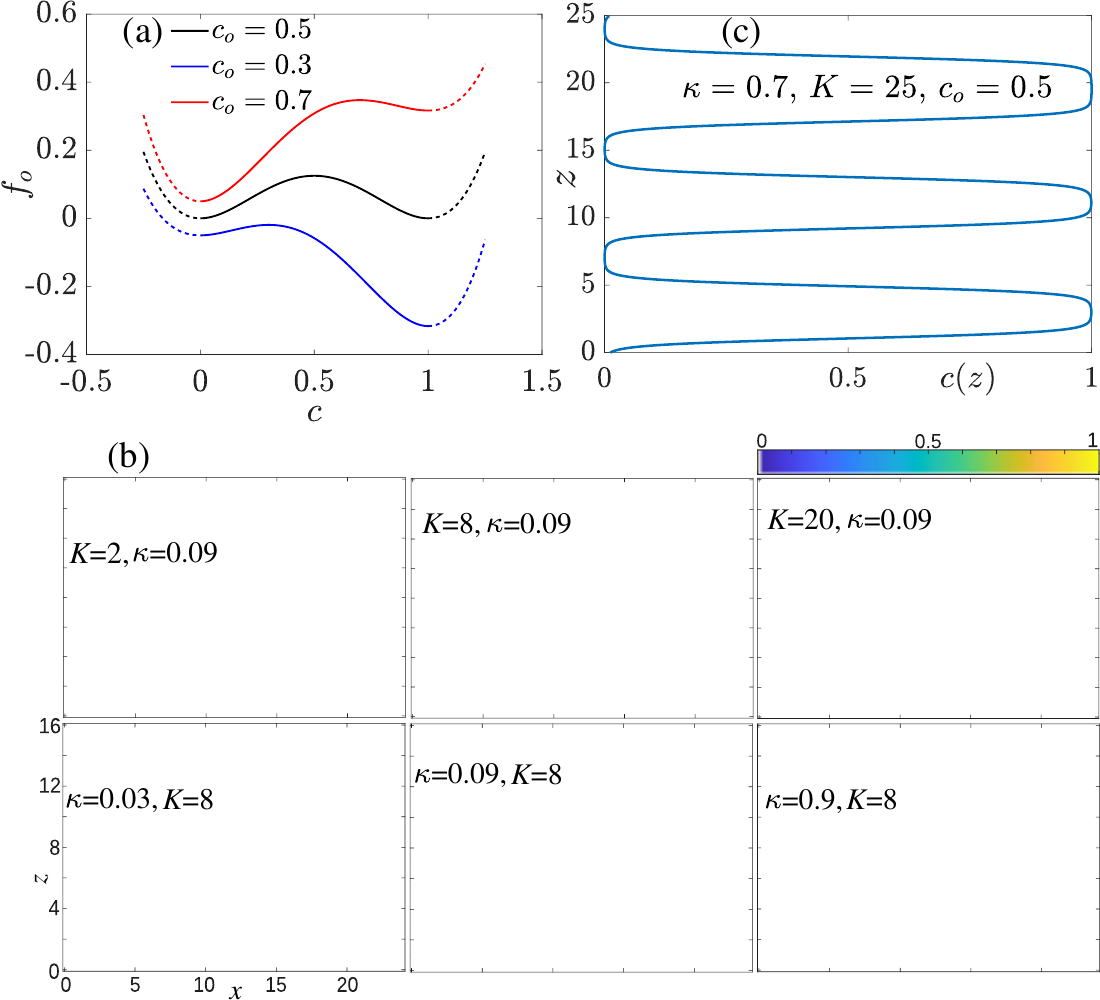}
	\caption{(a) The bulk free energy density $f_o$ used to construct the segregation flux [Eq.~\ref{eq_flux}], as a function of $c$ for different overall volume-fraction/concentration $c_o$. (b) Quasi steady-state solution of Eq.~\ref{eq_c} in 2D using the flux in Eq.~\ref{eq_flux} at different $K$ and $\kappa$ with $c_o=0.5$. (c) The steady-state solution of the 1D Eq. $0= M(\dot{\gamma}) \left[  A \partial^2_z c +B (\partial_z c)^2- \kappa \partial^4_z c \right]$, indicating that the layers are favored to achieve a balance between terms having $\partial_z^2c$ and $(\partial_z c)^2$ with non-linear coefficients, and the term having $\partial_z^4c$.}
	\label{fig_cahn-hilliard}
\end{figure}

\begin{figure}
	\centering
	\includegraphics[width=1.0\linewidth]{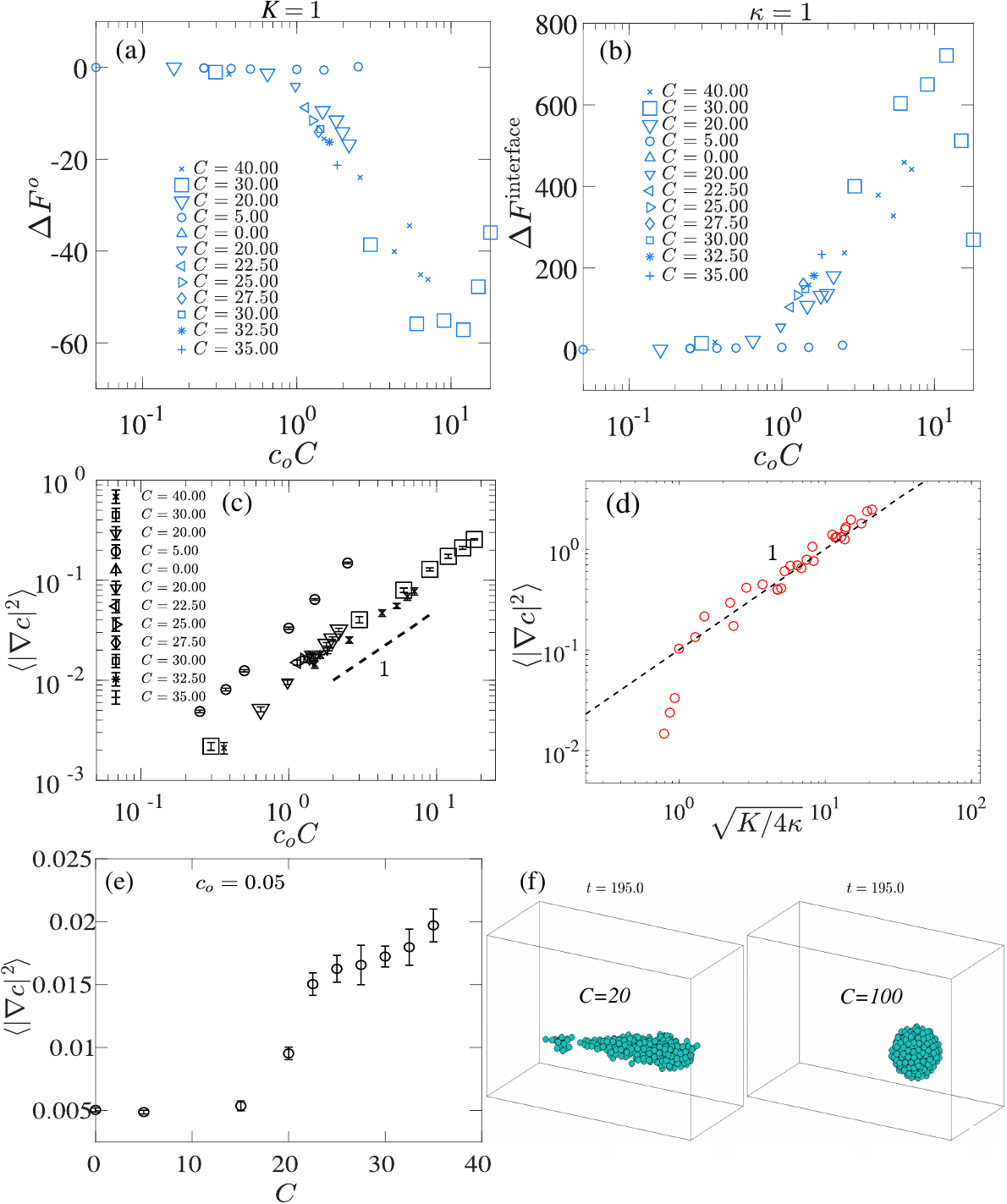}
	\caption{Changes in (a) the bulk free energy $F^o$ [Eq.~\ref{eq_Delta_F}], and (b) the interfacial free energy $F^\mathrm{interface}$ [Eq.~\ref{eq_Delta_F_interface}] as predicted from DEM simulations. (c) The average interfacial energy $\langle|\nabla c|^2\rangle$ as a function of the parameter $c_oC$ in the DEM simulations, and (d) as a function of the parameter $\sqrt{K/(4\kappa)}$ in the free energy based continuum simulations. Appearance of same scaling in both DEM as well as free energy based simulations establishes a correspondence between $c_oC$ and $\sqrt{K/(4\kappa)}$. (e) The interfacial energy as a function of $C$, indicating as if there exists a threshold $C$. (f, left) Breakup of a single cohesive agglomerate near threshold $C\approx20$; upon increasing $C$ to $\sim10^2$ (f, right) the agglomerate behaves almost like a solid object requiring consideration of complete momentum balance rather than imposition of simplified velocity profile $\mathbf{u} = \dot{\gamma} z  \mathbf{i}$.}
	\label{fig_interfacial_energy}
\end{figure}

Figure~\ref{fig_particles} shows quasi steady-state configurations for (a) relatively low $c_o$, low $C$; (b) low $c_o$, high $C$; and (c) high $c_o$, high $C$. By quasi steady-state, we imply that we are away from the initial transients in the system, but dynamic events such as intermittent agglomerate formation and breakup are permitted.
The cohesive grains remain homogeneous at low $c_o$ and low $C$ [Fig.~\ref{fig_particles} (a)]. At moderate $c_o$ and $C$, they form irregularly shaped lumps, as expected [Fig.~\ref{fig_particles} (b)]. However, upon increasing $c_o$ and $C$, we reach a point where the system segregates into percolating stripes or layers -- remarkably similar to the classical Cahn-Hilliard phase-separating system under shear [Fig.~\ref{fig_particles} (c)].
\begin{figure}
	\centering
	\includegraphics[width=1.0\linewidth]{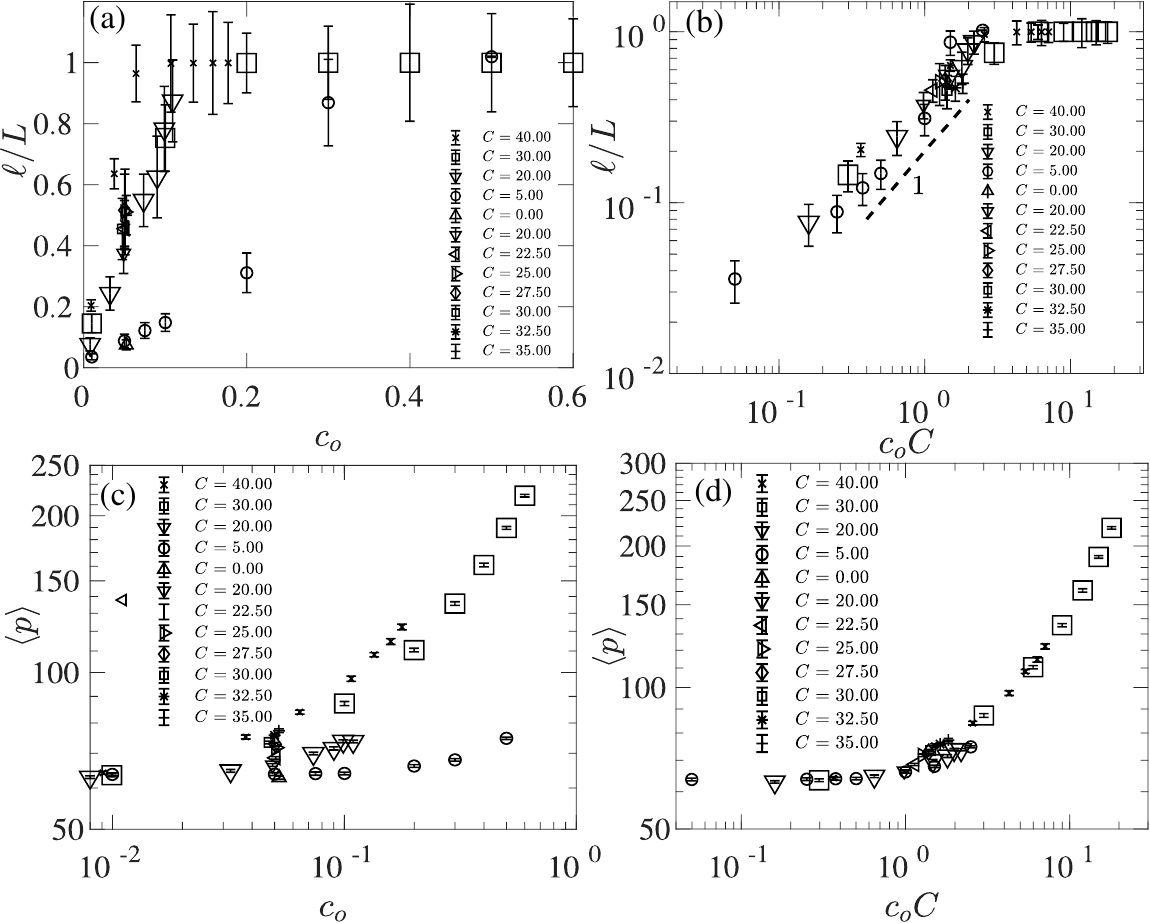}
	\caption{(a) Normalized linear size of clusters, $\ell/L$, as a function of the overall concentration of cohesive grains $c_o$. (b) The data for $\ell/L$ collapses onto a single curve $\ell/L\sim c_oC$ before the agglomerates show percolation ($\ell/L \approx 1$). (c) The space-time average of non-cohesive pressure is presented as a function of $c_o$, which also collapses onto a single curve when plotted against $c_oC$ (d). The error bars represent standard deviations over a quasi-steady-state time window.}
	\label{fig_scaling}
\end{figure}
A characteristic normalized linear size of the agglomerates, defined as 
$
{\ell}/{L} \equiv \max\left[ 
\langle {\ell_x}/{L_x} \rangle_t,\:
\langle {\ell_y}/{L_y} \rangle_t,\:
\langle {\ell_z}/{L_z} \rangle_t 
\right]
$, is shown in Fig.~\ref{fig_scaling} (a) as a function of $c_o$.
Here $\ell_k$ and $L_k$ are the agglomerate span and the flow bed size in the $k^\mathrm{th}$ direction, respectively, and $\langle\cdot\rangle_t$ implies time average over a quasi steady-state simulation period ($1500 \leq t \leq 2000$).
The data for $\ell/L$ collapses on a single curve when plotted against $c_oC$ [Fig.~\ref{fig_scaling} (b)]. We observe a power law scaling close to $\ell/L \sim (c_oC)^1$ for $c_oC<2$; for $c_oC>2$ we start to observe percolating agglomerates ($\ell/L \approx 1$). 
To obtain the concentration or the stress field, the microscopic DEM data is coarse-grained onto a mesoscopic grid [Fig~\ref{fig_particles} (d), Supplementary]. This allows us to obtain the average stress in the system from interparticle force data, and the average interfacial energy between cohesive and non-cohesive phases from the concentration field and its gradients.
The data for average normal stress or pressure also collapse onto a single curve when plotted against $c_oC$ [Fig.~\ref{fig_scaling} (c-d)]. 
The collapse and scaling in Fig.~\ref{fig_scaling} indicates that the dynamics are a function of the non-dimensional product $c_oC$.
The important question is to identify the macroscopic fluxes that lead to the observed dynamics and patterns, and relate them to the microscopic control parameter $c_oC$. 
We propose an ansatz that, in the present granular system, the development of interfaces between cohesive and non-cohesive components is akin to phase separation in binary molecular mixtures driven by an effective free energy, although we are dealing with a non-equilibrium system. 
The free energy, however, is to be activated by applying shear and do not lead to any motion or segregation on its own.
We enforce the conservation of concentration $c(\mathbf{r},t)$ of cohesive phase, and impose a linear macroscopic velocity profile, respectively 
\begin{align}\label{eq_c}
\partial_t c + \nabla \cdot (\mathbf{u}c) = -\nabla \cdot \boldsymbol{j} &= - \nabla \cdot \left( -M \nabla \frac{\delta F}{\delta c}\right),\\
\mathbf{u} = \dot{\gamma} z  \mathbf{i} &= \frac{v_\mathrm{top}}{L_z}  z  \mathbf{i},
\label{eq_velocity_profile}
\end{align}
where we assume that the non-advective flux $\boldsymbol{j}$ is derivable from an effective free energy functional $F[c]$. 
If there is no external driving, the granular system cannot spontaneously relax towards the minimum of this free energy. To enforce this, we posit that the mobility $M$ depends linearly on the strain rate, $i.e.$, $M(\dot{\gamma})\equiv M_o\dot{\gamma}$, leading to $\boldsymbol{j}=\mathbf{0}$ if $\dot{\gamma}=0$. In other words  $M(\dot{\gamma})$ sets the strength of the dynamics caused by $F[c]$. 
The macroscopic velocity $\mathbf{u}$ is in principle governed by the momentum balance, but for shear flow without gravity, we can safely impose Eq.~\ref{eq_velocity_profile} assuming low shear viscosity in the cohesive phase. 
Thus, constructing the non-advective flux $\boldsymbol{j}$ remains the central challenge~\citep{gray2018particle,umbanhowar2019modeling}.
Analogous to binary phase separation, we consider that
$
F[c] = \int f(c,\nabla c,\nabla^2 c, ...) d\mathbf{r}
$
where $f$ is the free energy density which depends on the concentration and its spatial gradients. A variation $\delta c(\mathbf{r})$ in the concentration field leads to a variation in the total free energy
$
\delta F = \int \frac{\delta f}{\delta c} \, \delta c \, d\mathbf{r}$, and the functional derivative $
\frac{\delta F}{\delta c}=\frac{\partial f}{\partial c} - \nabla\cdot\frac{\partial f}{\partial\nabla c}$ acts as a chemical potential. 
Further, $f$ can be decomposed as
$
f = f_o + \frac{1}{2}\kappa (\nabla c)^2
$
where $f_o$ is the bulk free energy density as if the system is in a homogeneous configuration, and $\frac{1}{2}\kappa (\nabla c)^2$ is the cost associated with concentration gradients or the interfaces~\citep{cahn1958free,cates2018theories}. 
To segregate the mixture, we force $f_o$ to have a double well shape: $f_o' =0$ at $c=0,1,c_o$; $f_o'' > 0$ at $c=0,1$; and $f_o'' < 0$ at $c=c_o$. The conditions are satisfied if we construct $f_o = K\left[ c_o c^2/2 -(1+c_o)c^3/3 + c^4/4 \right]$ [Fig.~\ref{fig_cahn-hilliard} (a)]. So we have
\begin{align} \label{eq_Fc}
 F[c] = \underbrace{ \! \int \!\! K\!\left[ c_o \frac{c^2}{2} -(1+c_o)\frac{c^3}{3} + \frac{c^4}{4} \right] \! d\mathbf{r} }_{F^o[c]} 
 + \underbrace{ \!\int\! \frac{1}{2}\kappa (\nabla c)^2 d\mathbf{r} }_{F^\mathrm{interface}[c]},
\end{align}
from which the flux $\boldsymbol{j} = -M(\dot{\gamma}) \nabla \frac{\delta F}{\delta c}$ comes out to be
\begin{equation}\label{eq_flux}
\boldsymbol{j} = M(\dot{\gamma}) \left[   -K[c_o -2c(1+c_o) + 3c^2] \nabla c + \kappa \nabla(\nabla^2 c) \right].
\end{equation}
Using this flux, the steady-state solution of Eq.~\ref{eq_c} is shown in Fig.~\ref{fig_cahn-hilliard} (b) for different $K$ and $\kappa$ at $c_o=0.5$, predicting the formation of layers or stripes. 
The 1D steady-state version of Eq.~\ref{eq_c} along the cross-stream direction $z$ reads
$
0=\partial_t c = M(\dot{\gamma}) \left[  A \partial^2_z c +B (\partial_z c)^2- \kappa \partial^4_z c \right],
$
where the coefficients $A = K[c_o-2c(1+c_o)+3c^2]$ and $B=K[6c-2(1+c_o)]$ depend on $c$. For $K=25$, $\kappa=0.7$, and $c_o=0.5$, the steady-state solution of 1D version is shown in Fig.~\ref{fig_cahn-hilliard} (c) which also reproduces layers of low and high $c$.
Thus, there is a signature that the terms with $\nabla^2c$ and $(\nabla c)^2$ having nonlinear coefficients, balance the term with $\nabla^4c$.
We test the free energy hypothesis by calculating the change in $F^o$ and $F^\mathrm{interface}$ in a given DEM run as
\begin{align}\label{eq_Delta_F}
\Delta F^o &= F^o[c(t)] - F^o[c(t')],\\ \label{eq_Delta_F_interface}
\Delta F^\mathrm{interface} &= F^\mathrm{interface}[c(t)] - F^\mathrm{interface}[c(t')],
\end{align}
where $t'$ is the time at the beginning when the system is well mixed, and $t$ is a time chosen when the system attains a quasi steady-state. The changes $\Delta F^o$ and $\Delta F^\mathrm{interface}$ with time are shown in Fig.~\ref{fig_interfacial_energy} (a, b). For $c_oC>1$ and for sufficient $c_o$, the DEM data in Fig.~\ref{fig_interfacial_energy} (a, b) clearly tells that there is a competition between bulk free energy minimization, and interfacial energy maximization.

A general question pertaining to our study is to establish a relation between $\kappa$ and $K$ in the free energy, and the parameters $c_o$ and $C$ in the microscopic DEM simulations.
The stripes grow out of initially excited modes of the concentration field -- if initially homogeneous $c_o$ is perturbed to $c_o+\delta c$ with $\delta c/c_o\ll 1$, then the linearized time evolution of $\delta c$ reads
$
\partial_t \delta c + \nabla \cdot (\mathbf{u} \delta c) = M(\dot{\gamma})\left[  K(c_o-2c_o(1+c_o)+3c_o^2)\nabla ^2 \delta c - \kappa \nabla^4 \delta c  \right]
$ which, upon transforming to Fourier space, provides
$
\partial_t \delta c_k = M(\dot{\gamma}) \left[ -K(c_o-2c_o(1+c_o)+3c_o^2)k^2 -\kappa k^4 - i\mathbf{k}\cdot \mathbf{u} \right] \delta c_k
$. Choosing cross-stream $\mathbf{k}$ $i.e.$ $\mathbf{k}\cdot \mathbf{u}=0$ and perturbing around say $c_o=0.5$ we get the critical wavenumber $k_c=\sqrt{K/(4\kappa)}$, which is approximately the number of layers excited in a system of size $2\pi$. 
We find a remarkable correspondence between the critical wavenumber $\sqrt{K/(4\kappa)}$ in the continuum approach and $c_oC$ in the DEM simulations. From the concentration field, we quantify the average interfacial energy, $\langle |\nabla c|^2 \rangle$, and find that it scales linearly with $c_oC$ in the DEM simulations, and also linearly with $\sqrt{K/(4\kappa)}$ in the continuum model [Fig.~\ref{fig_interfacial_energy} (c, d)]. The width of the layers $\sim 1/k_c$, and the size of the simulation box in $z$ has to be large enough to accommodate this. In Fig.~\ref{fig_interfacial_energy} (d), the data towards the lower end of $k_c$ drops sharply because in this range the number of waves excited in the system approach one.
Additionally, in DEM simulations we have observed that at a fixed concentration, the interfacial energy is negligible at low $C$ when cohesion is insufficient to form stable agglomerates, however, it exhibits a jump when $C$ is increased -- as if there exists a threshold $C$ -- for example for $c_o=0.05$ the interfacial energy jumps near $C\approx 20$ [Fig.~\ref{fig_interfacial_energy} (e)]. Absence of segregation at low $C$ can be replicated by choosing $K<0$ in the continuum model (and vice versa); $K<0$ allows only one minima in $f_o$, leading to a stable homogeneous concentration field. Thus, in addition to correspondence between $c_oC$ and $\sqrt{K/(4\kappa)}$, we can also say that $K$ in the continuum model is akin to $C-C_\mathrm{th}$ in the microscopic DEM model, where $C_\mathrm{th}$ is the threshold cohesive force parameter.

The above approach of construction of non-advective flux has helped us to predict steady-state layer formation under shear, as well as it is able to predict some transient phenomena such as pinchoff during breakup of a single cohesive agglomerate under shear [see Supplementary]. However, we observe limitations under stronger cohesion. When we increase $C$ to $\sim10^2$ keeping other parameters fixed, the aggregates that form remain intact [Fig.~\ref{fig_interfacial_energy} (f)]. In this case, the agglomerates behave almost like solid objects, and the counterpart of this in the continuum model is a large increase in shear viscosity of the cohesive phase. This feature cannot be incorporated unless one considers complete momentum balance with high viscosity difference between the two phases, instead of simplified velocity profile imposition. We leave this as an open question -- to relate $C$ with shear viscosity and to find out how it enters the stress-strain-rate relationship in a complete momentum balance.

In conclusion, we have simulated a dense mixture of cohesive and non-cohesive grains subjected to plane shear using DEM simulations, and report regular and irregular patterns forming due to cohesion-mediated segregation.
At moderate concentrations of the cohesive grains ($c_o$) and the strength of cohesive interactions ($C$), cohesive grains form irregular-shaped lumps as expected; however, upon increasing $c_o$ and $C$, we reach a point where the system segregates into percolating layers or stripes. 
We establish that the interfacial energy scales with $c_oC$ in the particle simulations and with $\sqrt{K/(4\kappa)}$ in the continuum model simulations.
This remarkable correspondence points us to a systematic way of constructing segregation fluxes in cohesive granular mixtures through an appropriately modified free energy principle. Our study opens a general question about whether the free energy principle can also be applied to model segregation driven by mechanisms other than cohesion -- the process which is traditionally thought to be distinct from equilibrium phase separation in molecular mixtures because of the presence of friction and inelasticity.

%
%
%
%
%\section*{Acknowledgements}
CS acknowledges support from the INSPIRE Faculty Fellowship (IVR No. 201900010281) of the Department of Science and Technology, India. This work started while CS was a researcher at Northwestern University, working with Paul Umbanhowar and Richard Lueptow, and CS acknowledges great benefit from a number of fruitful discussions with PU and RL.

%\bibliographystyle{apsrev4-2}
%\bibliography{ref}


\onecolumngrid
\clearpage
\section{Supplementary: Cohesion mediated layering in sheared grains}
\section{Setup of DEM simulations}
We use the discrete element method (DEM) to solve the equations of motion for grains. Initially, a total of $N=24\times24\times50$ particles are generated on a regular grid and are allowed to settle down on a plane bottom wall under gravitational force. For a simulation time $0\leq t\leq 200$ (in units of $\tau$ as explained below), the grains are allowed to execute gravity-driven chute flow by tilting the domain. This helps to initiate a dense granular flow. After $t=200$, the tilt angle is set to zero, and the particles within a layer at the top and within a layer at the bottom are flagged as topwall and bottomwall particles, respectively. The gravity is now turned off on all particles, except at the topwall particles as a way to apply pressure at the top. The topwall particles are now held together and moved with a constant speed $v_\mathrm{top}\mathbf{i}$ to apply the strain-rate, while the bottomwall particles are kept fixed by setting zero velocity.
We non-dimensionalize the system by measuring length, mass, and time in units of particle diameter $d$, particle mass $m$, and time scale $\tau= \sqrt{md/F}$, respectively, where $F$ is a force scale. We fix a gravity-based force scale $F=mg$ and thus $\tau=\sqrt{d/g}$, where $g$ is the gravitational acceleration. For shear flow, although we turn off the gravity for the particles flowing in the bulk, we still use a top wall made up of $N_\mathrm{top}$ particles on which gravity remains turned on as a way to apply pressure $P = N_\mathrm{top}/(L_x L_y)$ $[mg/d^2]$ at the top, where $L_x,L_y$ are the non-dimensional domain lengths in $x$ and $y$ direction respectively. We tune: ($i$) the {\bf overall volume-fraction/concentration} of the cohesive grains 
\begin{equation}
c_o = \frac{\sum_i d_i^3 \times \mathrm{flag}_i}{\sum_i d_i^3},
\label{eq_co}
\end{equation}
where $\mathrm{flag}_i=1$ for cohesive grains and $0$ for non-cohesive grains, and the sum is over the bulk flow particles, and ($ii$) a cohesive force parameter $C$. Instead of considering specific complexities of a cohesive interaction ({\it e.g.}, wetting or moisture absorption, due to surface treatment of particles with polymeric agents or chemical salinization, due to Van der Waals or electrostatic interactions), we use a simplified inter-particle cohesive force which increases linearly with particle overlap $\delta_{ij}$.
We define the {\bf cohesive force parameter} as
\begin{equation}
C = |\min[F_{ij}^\mathrm{elastic}+F_{ij}^\mathrm{cohesive}]|,
\label{eq_cohesive_force_parameter}
\end{equation}
where $F_{ij}^\mathrm{elastic}+F_{ij}^\mathrm{cohesive}$ is the sum of the normal elastic and cohesive forces in absence of normal damping, and $\min[F_{ij}^\mathrm{elastic}+F_{ij}^\mathrm{cohesive}]$ is its minimum value [Fig.~\ref{fig_force_model}]. 
This is essentially, in static balance, the net interparticle attractive force measured in units of $F=mg$; thus, $C$ is akin to the granular Bond number.
The size of the domain in $x,y$ is fixed to $24\times24$ $[d^2]$, while the flow depth of the particles in the bulk remains $\sim29$ $[d]$, the topwall speed is $v_\mathrm{top}=3$ $[d/\tau]$, the number of topwall and bottomwall particles are $\sim5600$ and $\sim2000$ respectively, and the particle packing fraction (volume occupied by the bulk flow particles/$L_xL_yL_z$) of particles remains $\sim0.61$ in all the simulations. Note that the bulk flow height $L_z\equiv L_z(t)$ as the topwall is permitted to adjust its height around a mean; the fluctuations in $L_z$ are small relative to its mean value.
Mathematical expressions of the cohesive and non-cohesive contact forces, torques, and model parameters are given below.

\section{Contact and cohesive interactions in DEM simulations}\label{app_a}
\begin{figure}
	\centering
	\includegraphics[width=0.65\linewidth]{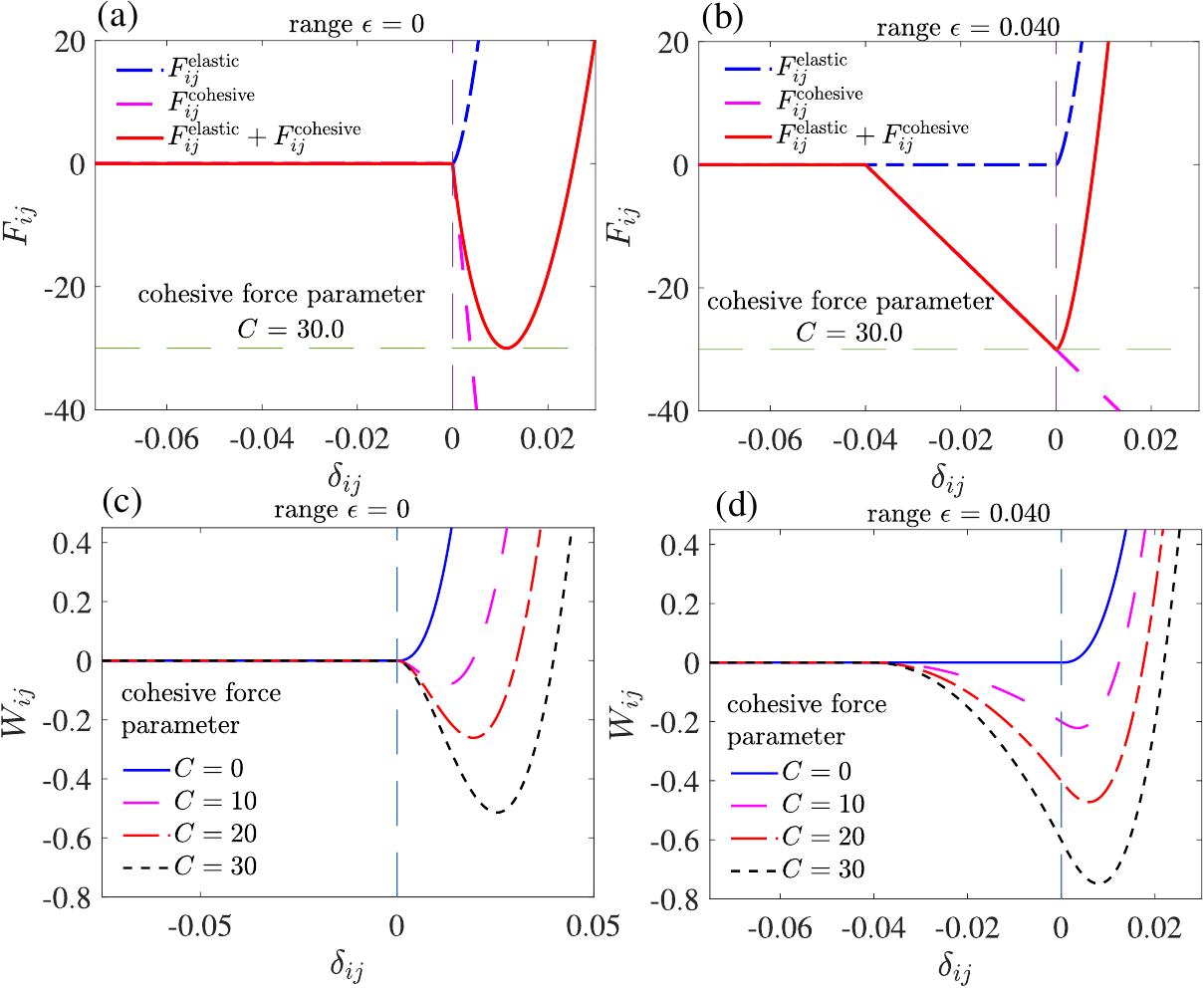}
	\caption{Cohesion model: Static balance between elastic and cohesive forces as a function of overlap between two particles for (a) model with zero range, and (b) model with a short range. (c, d) Corresponding potential curves $W_{ij}(\delta_{ij})=\int_{-\infty}^{\delta_{ij}}(F_{ij}^\mathrm{elastic}+F_{ij}^\mathrm{cohesive})\:d\delta_{ij}$ for different values of the cohesive force parameter $C$. The incorporation of a short range $\epsilon$ in the cohesion model does not cause any qualitative difference in the observed pattern formation.}
	\label{fig_force_model}
\end{figure}
A particle $i$ experiences a force exerted by a contacting particle $j$ normal to the plane of contact and acting in the direction from the center of particle $j$ towards the center of particle $i$, modeled as
\begin{equation}
\mathbf{F}_{ij}^\mathrm{n} =  \Theta(\delta_{ij})\:
(\delta_{ij} R_\mathrm{eff})^{1/2} \: 
\min \left[ 0,\: k_\mathrm{n} \delta_{ij}
- \gamma_\mathrm{n} \mathbf{v}_{ij}\cdot\mathbf{n}_{ij} \right] \:\mathbf{n}_{ij},
\end{equation}
where $\delta_{ij}=R_i+R_j-|\mathbf{r}_{ij}|, 
\mathbf{n}_{ij} =(\mathbf{r}_{i}-\mathbf{r}_{j})/
{|\mathbf{r}_{i}-\mathbf{r}_{j}|}$, $ \mathbf{v}_{ij} = \mathbf{v}_{i}-\mathbf{v}_{j}$, 
$R_\mathrm{eff}=R_iR_j/(R_i+R_j)$, $\Theta$ is the Heaviside step function, and $k_\mathrm{n},\:\gamma_\mathrm{n}$ are elastic and damping constants. We add a linear normal cohesive force
\begin{equation}
\mathbf{F}_{ij}^\mathrm{n,\:cohesive} =  \Theta(\delta_{ij})\:
\left[-c_\mathrm{n} \delta_{ij}\right]\:
\mathbf{n}_{ij},
\end{equation}    
where $c_\mathrm{n}$ is a cohesion constant. We set $c_\mathrm{n}$ as follows. In the absence of normal damping, the sum of the elastic and cohesive force is shown in Fig.~\ref{fig_force_model}. The extremum of attraction will occur when the particles are at an overlap $\delta_o$ such that
\begin{equation}
\frac{d}{d\delta_{ij}}[ F_{ij}^\mathrm{elastic}+F_{ij}^\mathrm{cohesive}]
=\frac{d}{d\delta_{ij}} \left[ (\delta_{ij} R_\mathrm{eff})^{1/2} k_\mathrm{n} \delta_{ij} - c_\mathrm{n} \delta_{ij} \right] = 0,
\end{equation}
which provides $\delta_o = \left[ \frac{2c_\mathrm{n}}{3 k_\mathrm{n} R_\mathrm{eff}^{1/2}} \right]^2$. At this $\delta_o$ the magnitude of attraction in Fig.~\ref{fig_force_model} is $|\min[F_{ij}^\mathrm{elastic}+F_{ij}^\mathrm{cohesive}]|$ which we call the {\bf cohesive force parameter} $C$. Putting $\delta_o$ we get $C=| \frac{-4c_\mathrm{n}^3}{ 27 k_\mathrm{n}^2 R_\mathrm{eff}}|$. 
Thus, in the simulations we set
\begin{equation}
c_\mathrm{n} = \left[ C \frac{27 k_\mathrm{n}^2 R_\mathrm{eff}}{4} \right]^{1/3},
\end{equation}
so that the maximum possible attractive force is $C$. Note that as we are measuring the forces in units of particle weight $mg$, the cohesive force parameter $C$ is akin to the granular Bond number.
\begin{figure}
	\centering
	\includegraphics[width=0.7\linewidth]{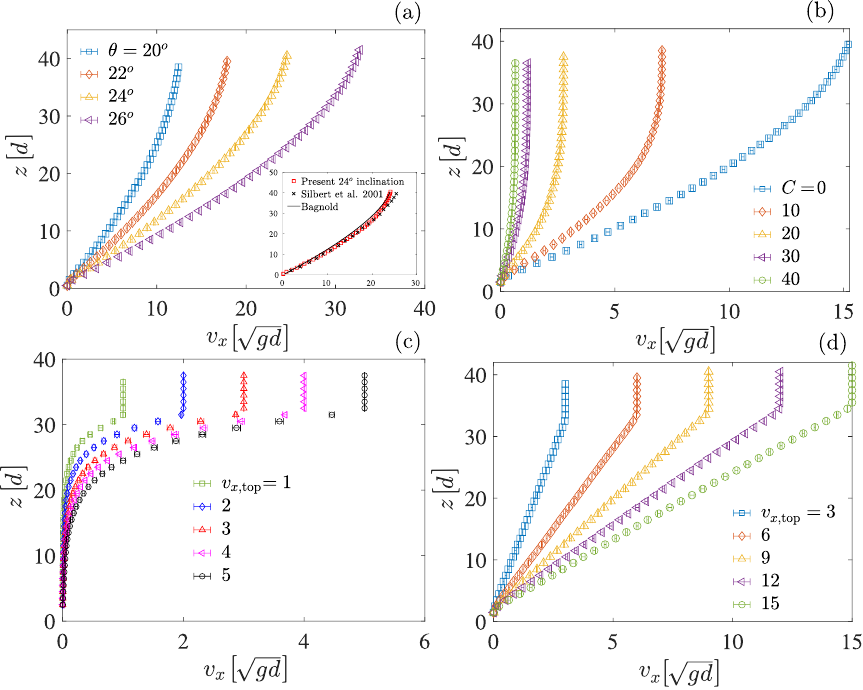}	
	\caption{Velocity profiles under different flow conditions: (a) free-surface inclined flow with gravity (and comparison with~\cite{silbert2001granular,bagnold1954experiments} in the inset); (b) free-surface inclined flow with gravity and cohesion (see $e.g.$~\cite{brewster2005plug}); (c) velocity and pressure at the top with gravity (see $e.g.$~\cite{kim2020power}); and (d) velocity and pressure at the top without gravity (see $e.g.$~\cite{kim2020power}). We have replicated these standard results while we have chosen the setup in (d) for the current study.}
\end{figure}
The tangential contact force is
\begin{equation}
\mathbf{F}_{ij}^\mathrm{t} = 
-\Theta(\delta_{ij})\:
(\delta_{ij}R_\mathrm{eff})^{1/2} \: 
\min\left[
\mu  |\mathbf{F}_{ij}^\mathrm{n}|,\:|k_\mathrm{t} \mathbf{s}_{ij}
+ \gamma_\mathrm{t} \mathbf{v}_{ij}^\mathrm{t}| \right]
\frac{\mathbf{s}_{ij}}{|\mathbf{s}_{ij}|},
\end{equation}
where
\begin{equation}
\mathbf{s}_{ij} = \int_{t_0}^{t} \mathbf{v}_{ij}^\mathrm{t}(\tau) \:d\tau
\end{equation}
is the accumulated tangential displacement starting at the beginning of a contact at $t_0$ and up to the current time $t$. If contact ends, the displacement is reset to zero. In practice, one has to rotate $\mathbf{s}_{ij}$ from the old time step so that it's always in the current tangential plane. For example, one could employ
\begin{equation}
\mathbf{s}_{ij}  
= 
|\mathbf{s}'_{ij}|
\frac{
	(\mathbf{s}'_{ij}\cdot\mathbf{v}_{ij}^\mathrm{t})}
{
	|\mathbf{s}'_{ij}\cdot\mathbf{v}_{ij}^\mathrm{t}|
}
\frac{\mathbf{v}_{ij}^\mathrm{t}}{|\mathbf{v}_{ij}^\mathrm{t}|}
+\mathbf{v}_{ij}^\mathrm{t}\Delta t,
\end{equation}
where $\mathbf{s}'_{ij}$ is the accumulated tangential spring displacement up to the previous time step, and $\mathbf{v}_{ij}^\mathrm{t}\Delta t$ is the addition in the current time step. Note that if $|\mathbf{v}_{ij}^\mathrm{t}|=0$ then the first term on the right side is singular and one has to come up with a better solution. In such a case, although there is no relative tangential displacement, the tangential plane may still have undergone rotation, or in other words, the particle pair in contact may have undergone rigid body rotation. The relative tangential velocity is
\begin{equation}
\mathbf{v}_{ij}^\mathrm{t} = \mathbf{v}_{ij} 
-  (\mathbf{v}_{ij}\cdot\mathbf{n}_{ij} ) \mathbf{n}_{ij}
- [{(R_i-\delta_{ij}/2)\boldsymbol{\omega}_{i}+(R_j-\delta_{ij}/2)\boldsymbol{\omega}_{j}}]\times \mathbf{n}_{ij}.
\end{equation}
The tangential forces contribute to torque on the particle and are given by
\begin{equation}
\mathbf{T}_{ij}^{\mathrm{t}}  =-\Theta(\delta_{ij})\:(R_i-\delta_{ij}/2)\:
\mathbf{n}_{ij}\times\mathbf{F}_{ij}^\mathrm{t}. 
\end{equation}
\begin{table}[b!]
	\caption{\label{tab:ex}Parameter values in DEM simulations.}
	\begin{ruledtabular}
		\begin{tabular}{lll}
			DEM simulation parameter & Description & Value \\	
			\hline	
			$L_x\times L_y\times L_z$  & Bulk flow bed size & $24\times24\times (29+\mathrm{fluctuation\:in\:bulk\:flow\:height})$\\
			$N,\:N_\mathrm{top},\:N_\mathrm{bottom}$ & Total, topwall, and bottomwall particles & $28800$, $\sim5600$, $\sim2000$\\				
			$d_i$  & Non-dimensional particle diameter & $0.95$ to $1.05$, uniformly distributed\\
			$m_i$  & Non-dimensional particle mass & $d_i^3$\\
			$v_\mathrm{top}$ or $v_{x,\mathrm{top}}$  & Topwall velocity & $3$\\							
			$k_\mathrm{n}$  & Normal stiffness & $10^5$\\
			$\gamma_\mathrm{n}$ & Normal damping  & $50$\\
			$k_\mathrm{t}$  & Tangential stiffness & $(2/7) k_\mathrm{n}$\\
			$\gamma_\mathrm{t}$  & Tangential damping & $1.0 \times 10^{-8}$\\
			$\mu$  & Tangential friction coefficient & $0.5$\\
			$\mu_\mathrm{r}$  & Rolling friction coefficient & $0.05$\\
			$c_o$  & Volume-fraction/concentration of cohesive grains & varied between $0$ and $0.6$\\
			$C$  & Cohesive force parameter & varied between $0$ and $40$\\										
		\end{tabular}
	\end{ruledtabular}
\end{table}
The rolling motion of particles is damped with a torque
\begin{equation}
\mathbf{T}_{ij}^{\mathrm{rolling}}  =
-\Theta(\delta_{ij})\:
(R_i-\delta_{ij}/2)\:
\mathbf{n}_{ij}\times
\mu_\mathrm{r}|\mathbf{F}_{ij}^\mathrm{n}|
\frac{[{(R_i-\delta_{ij}/2)\boldsymbol{\omega}_{i}-(R_j-\delta_{ij}/2)\boldsymbol{\omega}_{j}}]\times \mathbf{n}_{ij} }{|[{(R_i-\delta_{ij}/2)\boldsymbol{\omega}_{i}-(R_j-\delta_{ij}/2)\boldsymbol{\omega}_{j}}]\times \mathbf{n}_{ij} |},
\end{equation}
where $\mu_\mathrm{r}$ is the coefficient of rolling resistance and is set to a small value of $0.05$ in current simulations. Finally, using the above forces and torques, we solve the equations of motion
\begin{align}
m_i\frac{d\mathbf{v}_i}{dt} &= \sum_j ( \mathbf{F}_{ij}^\mathrm{n} + \mathbf{F}_{ij}^\mathrm{t} + \mathbf{F}_{ij}^\mathrm{n,cohesive} )\: \Theta(\delta_{ij}),\\
I_i\frac{d\boldsymbol{\omega}_i}{dt} &= \sum_j ( \mathbf{T}_{ij}^\mathrm{t} + \mathbf{T}_{ij}^\mathrm{rolling} )\:\Theta(\delta_{ij}),
\end{align}
where $I$ is the moment of inertia.
\begin{figure}
	\centering
	\includegraphics[width=0.65\linewidth]{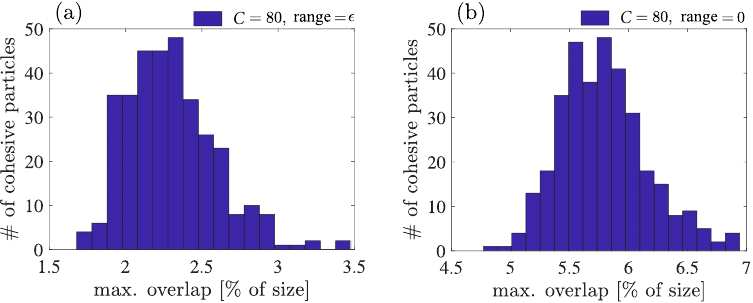}
	\caption{Maximum particle overlap in the limiting case of intact agglomerates (high $C$) at steady-state for a cohesive force model with (a) range $\epsilon=0.04$ [Fig~\ref{fig_force_model} (b, d)], and (b) range zero [Fig~\ref{fig_force_model} (a, c)]. Here $C=80$ and $v_{x,\mathrm{top}}=3$.}
\end{figure}

\section{Coarse-graining of DEM data onto a mesoscopic grid}
The space and time averages of concentration, velocity, and stresses are computed as follows. First, we divide the domain into a grid of cubic cells, and compute mesoscopic concentration and stress fields in a cell around $\mathbf{r}$ at $t$ as
\begin{align}\label{eq_c_meso}
c(\mathbf{r},t)  &\approx \frac{\sum_{i\in \Delta V} d_i^3 \times \mathrm{flag}_i} {\sum_{i\in  \Delta V} d_i^3},\\\label{eq_sigma}	
\sigma_{\alpha\beta}(\mathbf{r},t) &\approx -\frac{1}{\Delta V}\sum_{i\in \Delta V} \left [\sum_j r_{ij}^\alpha F_{ij}^\beta \: \Theta (R_i+R_j-r_{ij})\right],		
\end{align}
%\mathbf{u}(\mathbf{r},t) &\approx \frac{\sum_{i\in \Delta V} \mathbf{v}_i} {N_{\Delta V}},\\
%
%
%
%
where $\Delta V$ is the volume of the cell, $\mathrm{flag}_i=1$ for cohesive grains and $0$ for non-cohesive grains, $\Theta$ is the step function, and $R$ is grain radius. We can set $F_{ij}=F_{ij}^\mathrm{n,cohesive}$ to get the stresses due to cohesion only, or $F_{ij}=F_{ij}^\mathrm{n}+F_{ij}^\mathrm{t}$ to get the stresses due to elasticity plus friction/damping only. Thus, we can have a separate description for stresses due to cohesive and non-cohesive effects.
Next, we define the space and time averaged value of a mesoscopic field $\phi(\mathbf{r},t)$ as
\begin{align}\label{eq_phi}
\langle \phi \rangle &= \frac{1}{T} \int dt  \left[  \frac{1}{V} \int d\mathbf{r}\:  \phi(\mathbf{r},t)  \right] \approx \frac{1}{N_t}\sum_{n=1}^{N_t}  \frac{1}{N_\mathrm{cells}}\sum_{k=1}^{N_\mathrm{cells}} \phi_{k,n} ,
\end{align}
where $\phi_{k,n}$ is $\phi$ in $k^\mathrm{th}$ mesoscopic cell at $n^\mathrm{th}$ observation time instant, $V$ is the volume of the entire system, $T$ is the observation time period in a quasi steady-state, $N_\mathrm{cells}$ total number of mesoscopic cells, and $N_t$ are the total number of observation time instants in $T$. For example, average pressure would be $\langle p \rangle = \frac{1}{N_t}\sum_{n=1}^{N_t}  \frac{1}{N_\mathrm{cells}}\sum_{k=1}^{N_\mathrm{cells}} (-1/3)(\sigma_{xx} + \sigma_{yy} +\sigma_{zz})_{k,n} $, where $\sigma_{ii}$ are calculated using Eq.~\ref{eq_sigma}. We chose the observation time window $1500 \leq t \leq 2000$. The mesoscopic cells are of size $\Delta V = 2\times2\times2$ in units of $d\times d\times d$, however, we further use linear interpolation to smooth the fields for better resolution.

\section{Choice of bulk free energy density}
\begin{figure}
	\centering
	\includegraphics[width=0.7\linewidth]{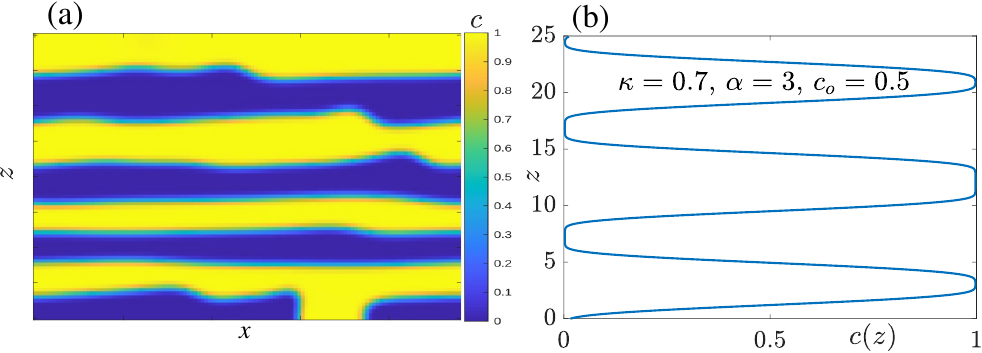}
	\caption{(a) Quasi-steady-state concentration field in 2D using the flux in Eq.~\ref{eq_flux_log}, with $\kappa=0.09$, $\alpha=3$. (b) Quasi-steady-state concentration field in the corresponding 1D version.}
	\label{fig_cahn-hilliard-log}
\end{figure}
The segregation is triggered using a double well shaped bulk free energy density, and for this, there can be multiple choices. For example the function $f_o =  c(1 - c) + \frac{1}{2\alpha} \left[ c \log c + (1 - c) \log(1 - c) \right]$ which is typically used in binary molecular phase separation, can also be used. However we can compare the parameters between different models by deriving the critical wavenumber as follows. The form $f_o =  c(1 - c) + \frac{1}{2\alpha} \left[ c \log c + (1 - c) \log(1 - c) \right]$ assumes that it centers around $c_o=0.5$. It implies
$
\frac{\delta F}{\delta c} = 1 - 2c + \frac{1}{2\alpha} \log \left( \frac{c}{1 - c} \right) - \kappa \nabla^2 c
$, and the flux comes out to be
\begin{equation}\label{eq_flux_log}
\boldsymbol{j} = M \left[   2\nabla c + \kappa \nabla(\nabla^2 c) - \frac{1}{2\alpha} \nabla \log \frac{c}{1 - c} \right].
\end{equation}
So here in comparison to Eq.~\ref{eq_flux}, the coefficient in front of $\nabla c$ is constant, but the flux has an additional logarithmic term. A typical quasi-steady state solution in 2D using this form of the flux is provided in Fig.~\ref{fig_cahn-hilliard-log} (a). Again, if we confine ourselves with variation of $c$ along $z$ only neglecting its variations along $x$ and $y$, then in steady state the equation reduces to
$
0=\partial_t c =  - M \left[  2 \partial^2_z c + \kappa \partial^4_z c - \frac{1}{2\alpha}  \partial^2_z \log  \frac{c}{1 - c}   \right],
$
whose solution agrees with the formation of layers of high and low concentration [Fig.~\ref{fig_cahn-hilliard-log} (b)]. 
How does the parameter $\alpha$ in this model compare with the parameter $K$ in the model in the main text?
For this, we look into the critical wavenumber. If an initially homogeneous $c_o$ is perturbed to $c_o+\delta c$ where $\delta c/c_o\ll 1$, then the evolution of $\delta c$ obeys 
$
\partial_t \delta c + \nabla \cdot (\mathbf{u} \delta c) = -M\left[ 2\nabla ^2 \delta c + \kappa \nabla ^4 \delta c - \frac{1}{2\alpha} \nabla^2 \log \frac{c_o+\delta c}{1 - c_o-\delta c} \right]
$, 
which,h upon linearizing and transforming to Fourier space, provides
$
\partial_t \delta c_k + i\mathbf{k} \cdot \mathbf{u} \delta c_k = -M k^2 \left[  \kappa k^2 - \left(2-\frac{1}{2\alpha c_o(1-c_o)}\right) \right] \delta c_k
$. Choosing cross-stream $\mathbf{k}$ $i.e.$ $\mathbf{k}\cdot \mathbf{u}=0$ and perturbing around say $c_o=0.5$ we get conditions $k^2 \leq \frac{1}{\kappa}\left(2-\frac{1}{2\alpha c_o(1-c_o)}\right)$ and $ \alpha\geq 1$ for the system to be unstable. For $c_o=0.5$, the critical wavenumber is $k_c = \sqrt{\frac{2}{\kappa}\left(1-\frac{1}{\alpha}\right)}$. Thus $2\left(1-\frac{1}{\alpha}\right)$ is akin to $K/4$ in the polynomial model constructed in Eq.~\ref{eq_Fc} in the main text. Numerically, Eq.~\ref{eq_Fc} in the main text has advantage that as $c\rightarrow 0$, there is no numerical blow up while Eq.~\ref{eq_flux_log} need special treatment as $c\rightarrow 0$.

\section{Application to single agglomerate breakup}
\begin{figure}[h!]
	\centering
	\includegraphics[width=1.0\linewidth]{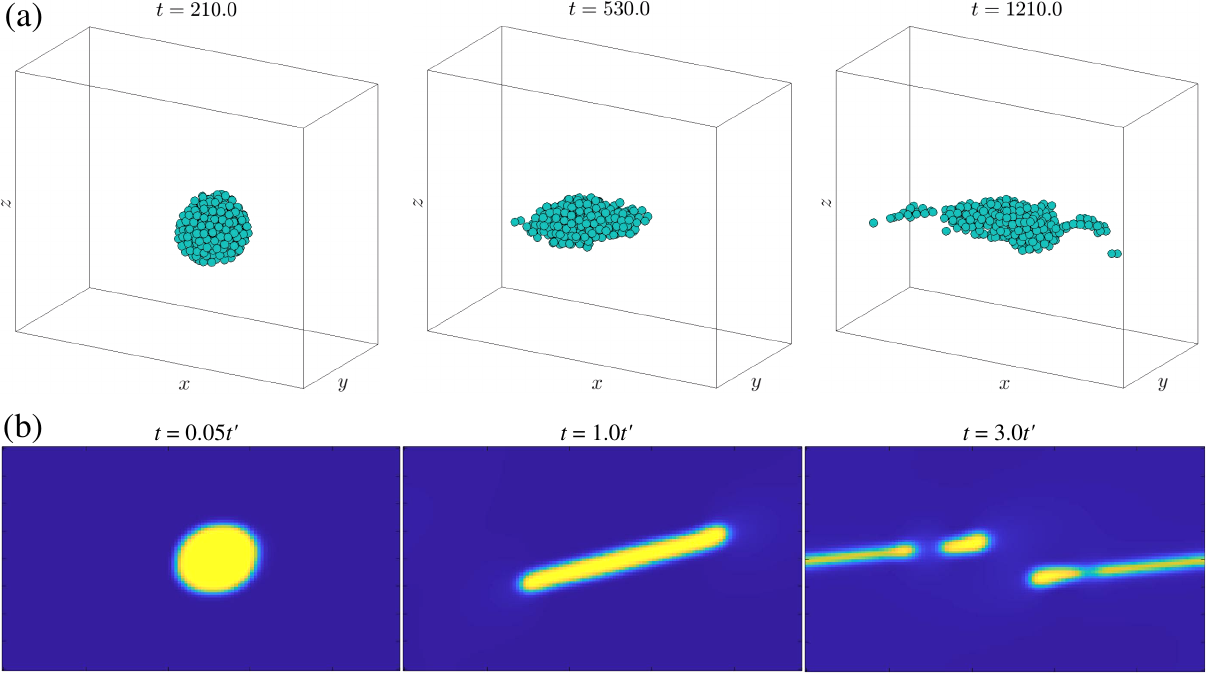}
	\caption{Breakup of a single agglomerate under shear from (a) DEM simulations, and (b) from continuum model with the flux constructed using free energy principle (main text). For moderate $C$, the continuum model predicts qualitative features such as pinchoff at the front and rear of the agglomerate. However, the cohesive agglomerate behaves almost like a solid object upon increasing $C$ -- as shown in the main text -- requiring consideration of the complete momentum balance rather than imposition of a simplified velocity profile $\mathbf{u} = \dot{\gamma} z  \mathbf{i}$.}
	\label{fig_blob}
\end{figure}

We test whether our constructed fluxes also apply to the limiting case of a single agglomerate under shear. Figure~\ref{fig_blob} (a) shows the DEM simulation of agglomerate stretching and then breaking at the front and as well as at the rear. Qualitatively, we observe similar behavior of the concentration field in the continuum model [Fig.~\ref{fig_blob} (b)]. The difference is that the cohesive agglomerate behaves almost like a solid object upon increasing $C$ -- as shown in the main text -- limiting the applicability of our continuum model. At increased $C$, the model should replicate a large increase in viscosity, which requires consideration ofthe  complete momentum balance equation rather than the imposition of a simplified velocity profile $
\mathbf{u} =  \frac{v_\mathrm{top}}{L_z}  z  \mathbf{i} = \dot{\gamma} z  \mathbf{i}$.

\clearpage
\section{Typical grain configurations}
\begin{figure}[h!]
	\centering
	\includegraphics[width=1.00\linewidth]{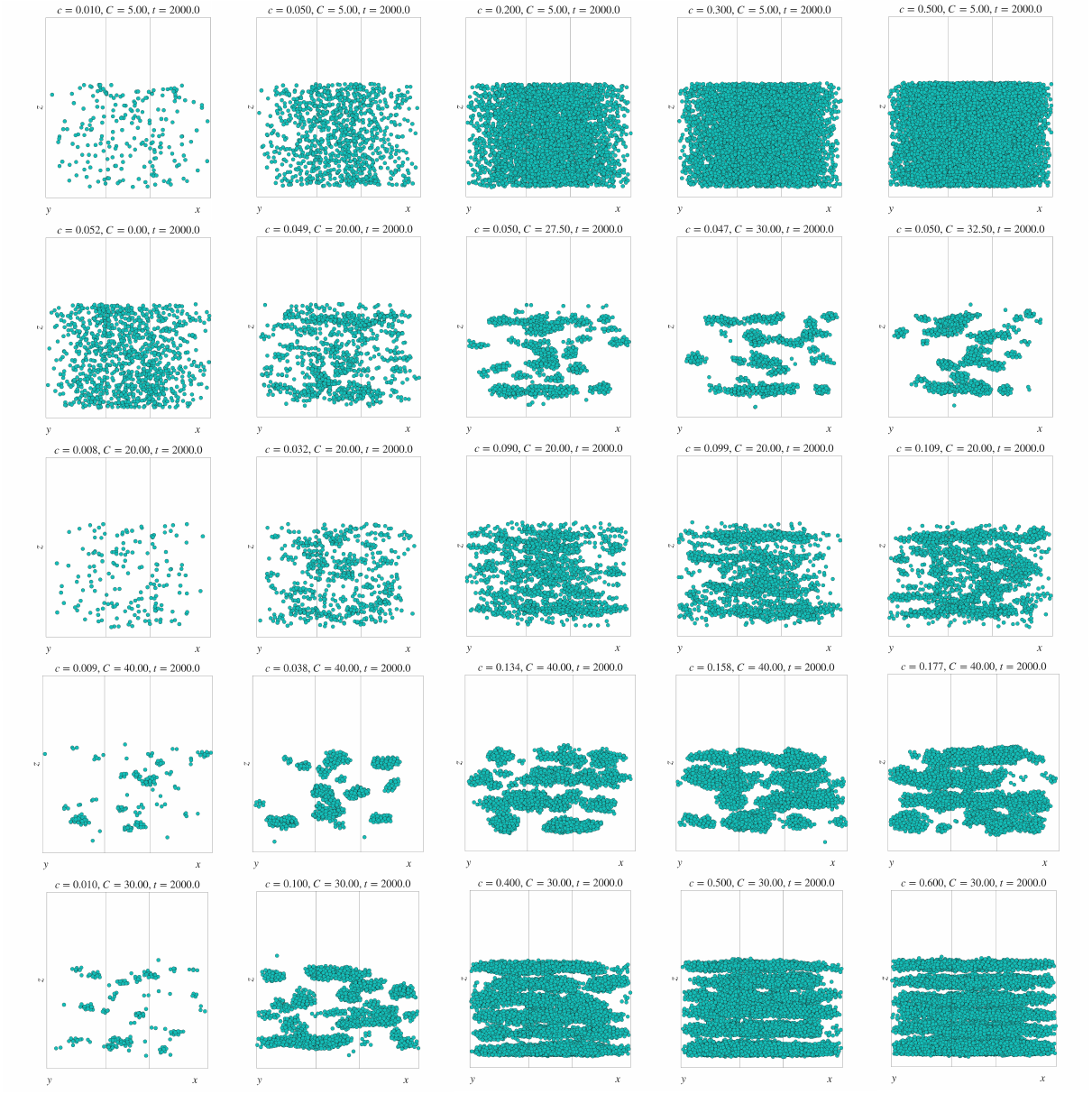}
	\caption{Typical grain configurations from DEM simulations at quasi-steady state and at different $c_o$ and $C$.}
	\label{fig_grain-configurations}
\end{figure}

%\bibliographystyle{apsrev4-2}
%\bibliography{ref}

\end{document}